\documentclass{ws-procs9x6}

\begin{document}

\title{TRACK FINDING IN GAMMA CONVERSIONS IN CMS}

\author{N. MARINELLI}

\address{Physics Department, University of Notre Dame,\\
Notre Dame, Indiana, USA\\
$^*$E-mail: nancy.marinelli@cern.ch}

\begin{abstract}
 A track finding algorithm has been developed for
 reconstruction of e$^+$e$^-$ pairs. It combines
 the information of the electromagnetic calorimeter with
 the information provided by the Tracker.
 Results on reconstruction efficiency of
 converted photons, as well as on fake rate are shown for single isolated photons and for photons from
 H$\rightarrow \gamma\gamma$ events with pile-up events at \mbox{10$^{33}$ cm$^{-2}$ s$^{-1}$} LHC luminosity.

\end{abstract}


\bodymatter

\section{Introduction}

The need for high granularity and an adequate number of measurements  
along the charged particle trajectories, in order to obtain 
excellent momentum resolution and pattern recognition in the
congested environment of LHC events, has lead to a CMS Tracker design having 
an unprecedented  large area of silicon detectors with a very large number of  front-end readout channels.
The resulting amount of material is large and comprises active layers, support structures, general services as well as 
an impressive cooling system.

\begin{figure}[h!]
\begin{center}
\vspace{5.8cm}
\includegraphics{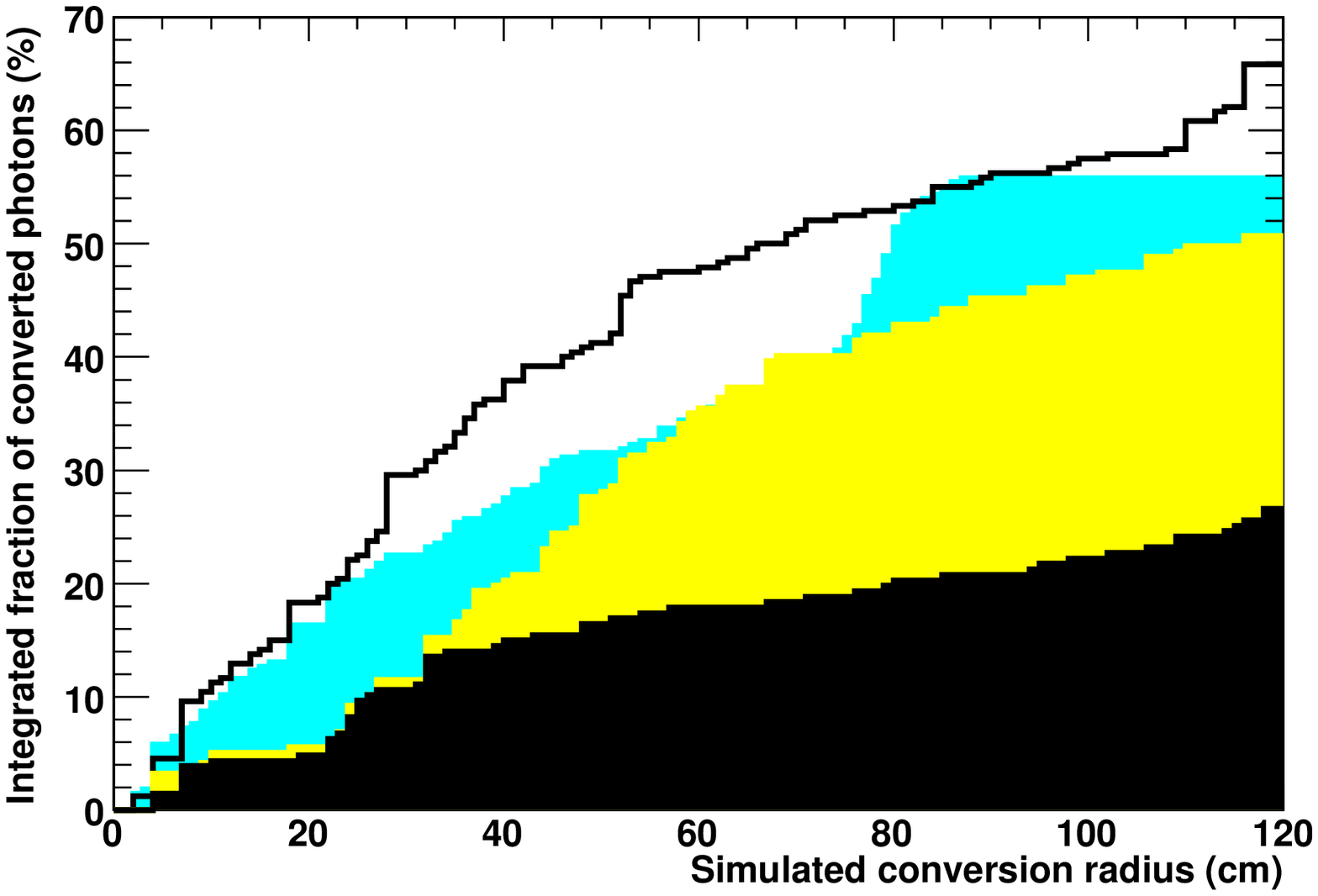}

\end{center}
\vspace{-2cm}
\caption{\small {
    Fraction of photons  converting in the Tracker, integrated over all radii. 
    The four histograms correspond to 0.1 slices in 
    $\eta$ around $|\eta|=0.2$ (black), $|\eta|=0.9$ (light grey), $|\eta|=2.0$ (dark grey)    and $|\eta|=1.2$ (hollow). } }
\label{Fig1}
\end{figure}

The relatively massive Tracker results in a large probability of photon conversion and electron bremsstrahlung radiation
in the Tracker volume. The fraction of photons converting in the Tracker material, integrated over the acceptance,
has been estimated from a simulated sample of single photons with P$_T$=35 GeV (Fig.~\ref{Fig1}). 
The number of gamma conversions in CMS is not negligible 
and it is important to reconstruct the e$^+$e$^-$ tracks.
Major examples of the use of track reconstruction of gamma conversions are:


$\bullet$ Photons from neutral pion decays constitute a very large background to prompt photons. In the case of
converted photons the rejection of the background using  the electromagnetic shower shape 
becomes ineffective. The information added by dedicated track finding improves the available rejection factor.
Moreover the information from the tracks can be used to
refine the electromagnetic energy clustering in the ECAL hence improving the energy measurement.


$\bullet$ Gamma conversion reconstruction is also a tool for electron reconstruction validation, i.e. asymmetric conversions
occurring very early in the Tracker are unwanted background to genuine electrons.

$\bullet$ The reconstruction of conversion vertices provides, once the reconstruction
efficiency is taken into account,  a ``radiography'' of the 
Tracker and allows the mapping of the material with data.

This paper summarizes the work described in Ref.~\cite{myNote,PTDR1}.  
A detailed description of the CMS Electromagnetic Calorimeter
and of the Pixel and Silicon Strip Tracker is provided in Ref.~\cite{ECAL_TDR} and Ref.~\cite{TRK_TDR,TRK_TDR_ADD} respectively.

\section{Electron-positron pair track reconstruction}
CMS track reconstruction is divided into four separate steps;
a) trajectory seed generation; b) trajectory building (i.e. seeded pattern recognition);
c) trajectory cleaning which resolves ambiguities and d) trajectory smoothing (i.e. the 
final track fit).

In CMS,  the standard seed and track finding algorithm~(Chapter 14, Sec.4.1 of Ref.~\cite{HLT_TDR}) 
was developed and optimized for tracks originating from the primary interaction vertex, 
with pattern recognition starting from track seeds built in the pixel detector. 
For electron tracks, instead, the match between a  super-cluster energy deposit in the ECAL 
and hits in the  pixel detector is used for building seeds (Ref.~\cite{standard_electrons}).

Neither of these approaches are  suitable for tracks originating from vertices   
that are significantly displaced with respect to the primary vertex such as those from  converted photons.
Different seed finding and pattern recognition algorithms are necessary.

Recently, after a major re-working of the CMS Reconstruction software took place (Ref.~\cite{CMSSW}), 
additional track seed finding methods were developed which no longer rely on the Pixel information.
However they were developed for general usage and do not
combine information from ECAL for specific conversion reconstruction. 

This paper describes the combination of an inward ECAL seeded track finding method with a 
subsequent outward track finding step.

\subsection{Inward Tracking}
\label{section_OI_tracking}
The electron bending in the CMS \mbox{4 Tesla} magnetic field and 
the large emission probability of bremsstrahlung photons in the Tracker material leads
to a spray of energy in the ECAL extending mainly in the transverse plane.
When dealing with single, high-energy  electrons the electron energy is collected by clusters
of clusters extended along a $\phi$ road called super-clusters (SC). Different clustering 
algorithms (Ref.~\cite{standard_electrons}) are used for the ECAL barrel and endcaps. 
The same clustering procedure is applied here when dealing with converted photons.

The initial assumption is made that the bulk of energy arising from converted photons is 
contained in one super-cluster, however allowance is made for tracks falling outside its boundaries.
The energy of the sub-clusters within a SC and the magnetic field
are used to give a first rough estimate of a trajectory path, assuming
that the initial photon vertex is the origin of the reference frame. 
Compatible hits are then sought for in the three outermost layers of Tracker. 
If compatible hits are found they are
used to re-evaluate the seed parameters releasing the initial hypothesis on the initial vertex.
Seeds are built out of pairs of hits and used for pattern recognition, 
trajectory building and final fitting proceeding inward in the Tracker, using the Kalman Filter 
formalism (Ref.~\cite{KalmanFilter}).

The average radiation energy loss (bremsstrahlung) experienced by electrons traversing 
the Tracker material is described by the Bethe-Heitler parametrization (Ref.~\cite{BetheHeitler,FittingWithEnergyLoss}). 
With the Kalman filter (which is a linear least-squares estimator),  the radiation energy 
loss is taken into account at each propagation step by correcting the track momentum by an amount corresponding to 
the predicted mean value of the energy loss and by increasing   
the track momentum variance with the predicted variance of the energy 
loss under the assumption that its distribution is Gaussian.

\subsection{Outward Tracking}
\label{section_IO_tracking}
The two oppositely charged tracks with the largest number of hits reconstructed with the inward tracking are   
used in turn, independently from one another,  as the basis for the outward seed and track finding procedure.
If only one track was found it is used by default. 
Given an inward track, its innermost hit is assumed to be the e$^+$e$^-$ vertex and  is used
as the starting point for seed finding of the other  conversion arm.

The first hypothesis of the outgoing track is made based on the presence of a basic cluster within a suitable
$\phi$ range from the presumed conversion vertex and the fact that the two tracks must be parallel at the vertex.
Pairs of hits compatible with the estimated track path 
are sought in the next two layers moving outwards along the helix.
These pairs are used as seeds for the forward trajectory building.
After this step, trajectories are cleaned according to the number of shared hits 
and smoothed with the backward fit  to obtain the parameters
of the tracks at their innermost state.
The description given in Sec.~\ref{section_OI_tracking} concerning the treatment of radiation energy losses  
applies here.

The two sets of tracks reconstructed in the inward and outward tracking procedures are
merged together.
All combinations of oppositely charged tracks are fitted to a common vertex 
and are considered as possible converted photon candidates. 
\subsection{Results}
\begin{figure}[h!]
\begin{center}
\vspace{6cm}
\includegraphics{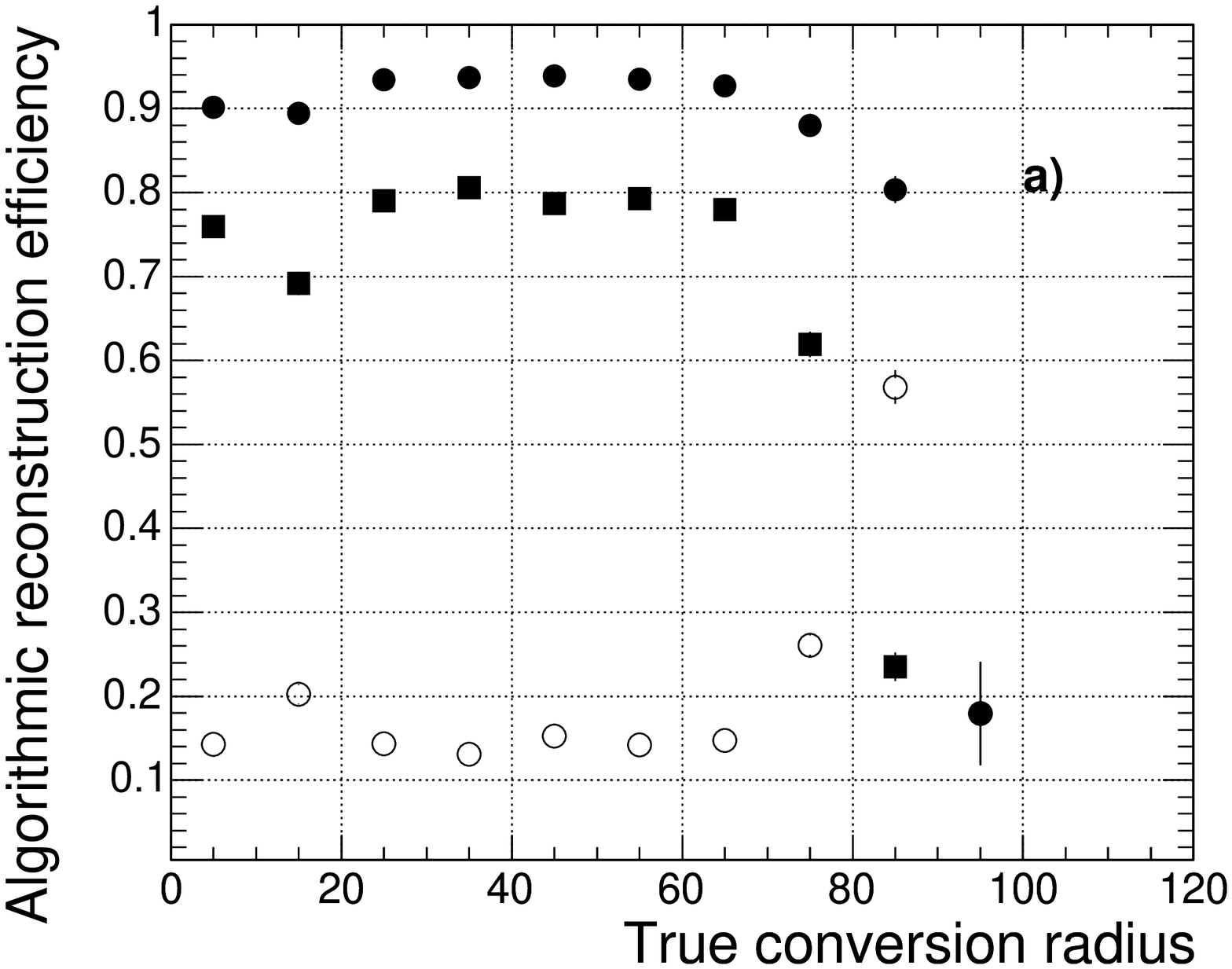}
\includegraphics{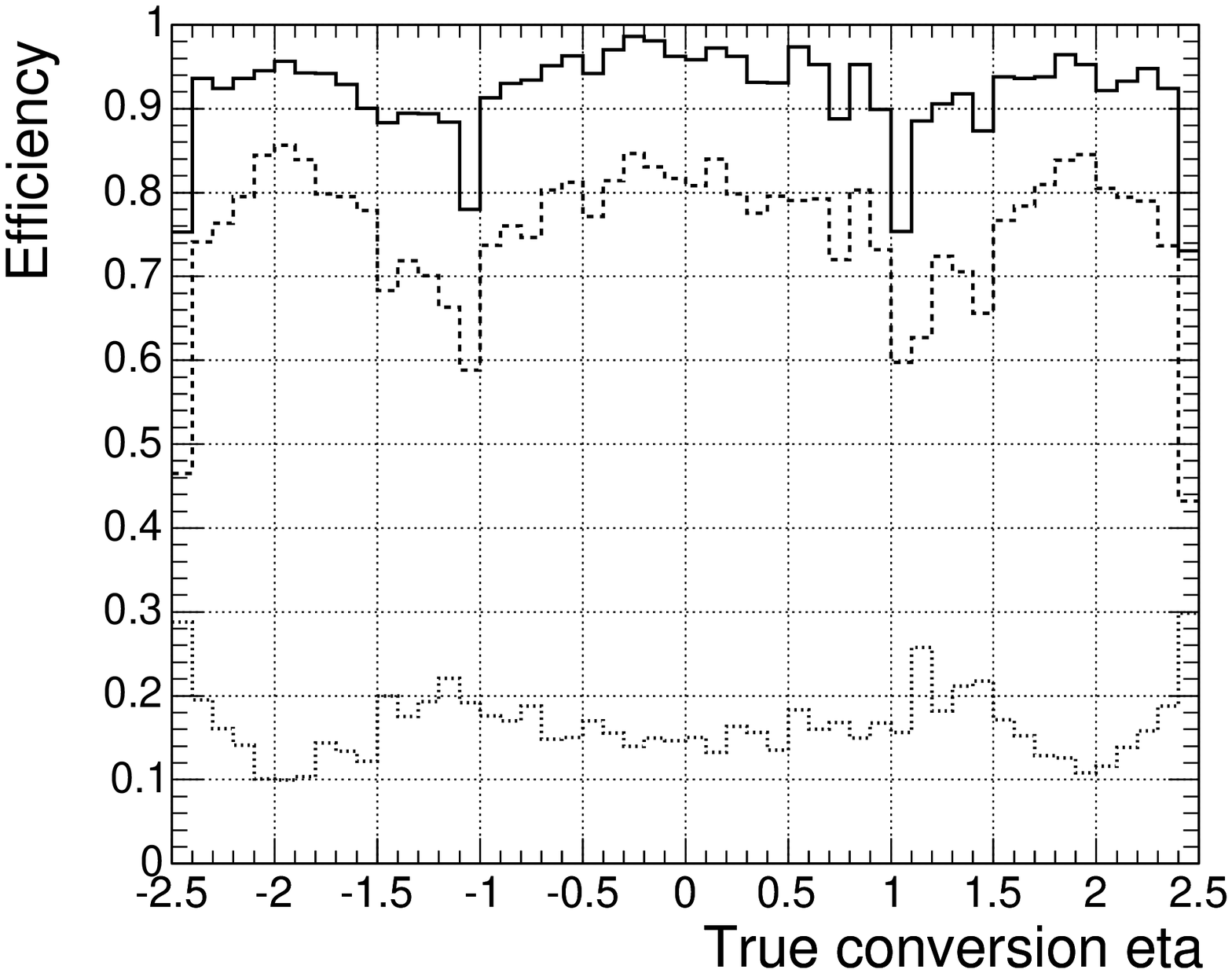}

\end{center}
\vspace{-2cm}
\caption{\small {
Reconstruction efficiency measured with single photons with 
fixed $P_T$=35 GeV/$c$ as a function of the simulated conversion-point position. 
(left) Total (black dots), two-tracks (black squares) and single track (open dots); 
(right) Total (solid line), two-tracks (dashed line) and single track (dotted line).  }}
\label{Fig2}
\end{figure}

The algorithmic efficiency was measured normalizing the number of reconstructed conversions to the simulated
conversions with the vertex located before the third-outermost Tracker layer (R$\sim$ 85 cm).
Figure~\ref{Fig2} shows the efficiency as a function of the radius and of $\eta$; the total efficiency
is broken down into two contributions arising from candidates with two reconstructed tracks and
those with only one track.

\begin{figure}[h!]
\begin{center}
\vspace{6cm}

\includegraphics{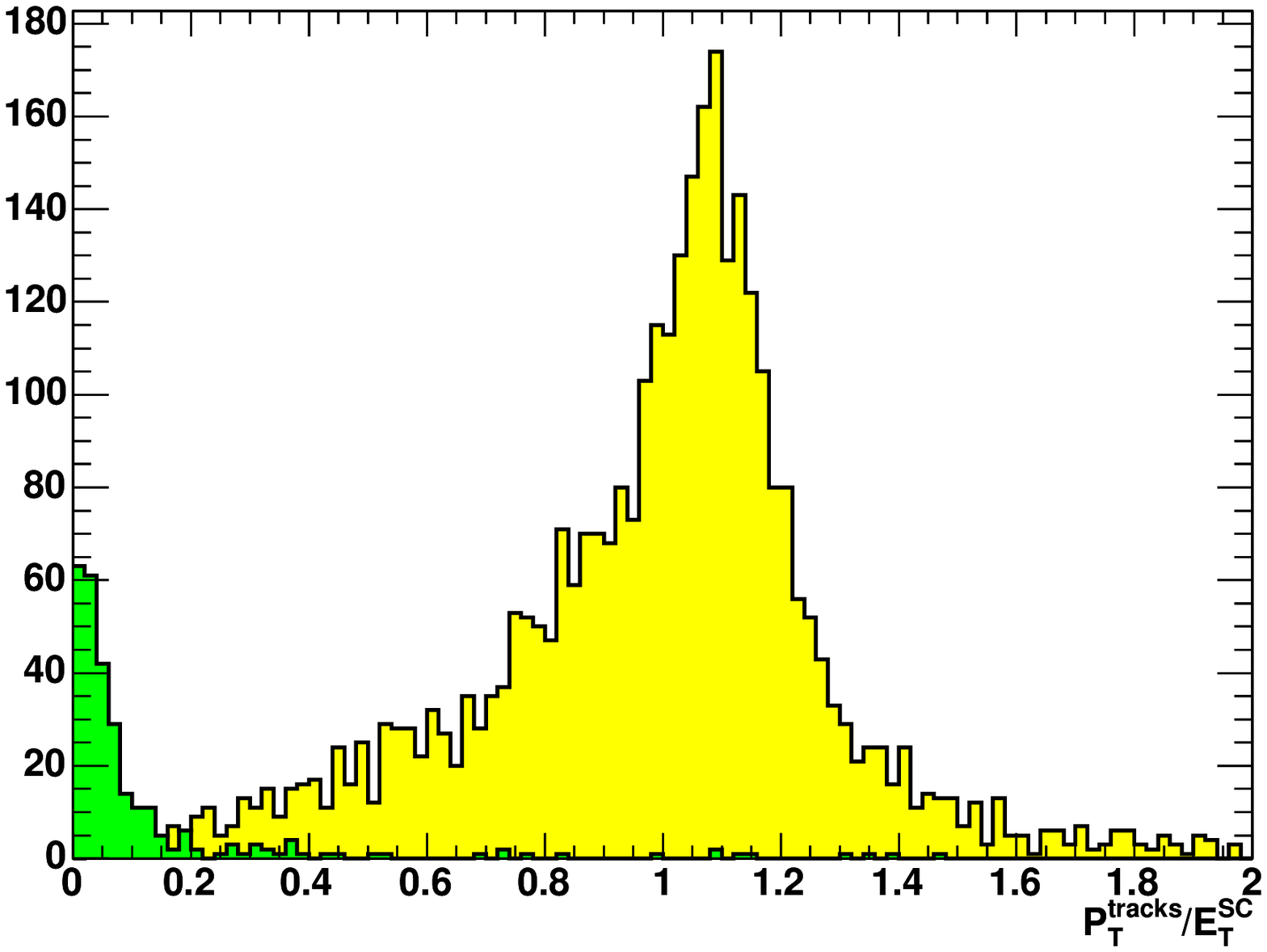}

\end{center}
\vspace{-2cm}

\caption{\small {The ratio $P_T({\mathrm {tracks}})/E_T({\mathrm {SC}})$  
for reconstructed converted photons in a sample  of H$\rightarrow \gamma\gamma$ events with low luminosity pile-up.
The dark grey histogram shows the contribution from fake pairs. }}
\label{Fig3}
\end{figure}

\begin{figure}[h!]
\begin{center}
\vspace{6cm}
\includegraphics{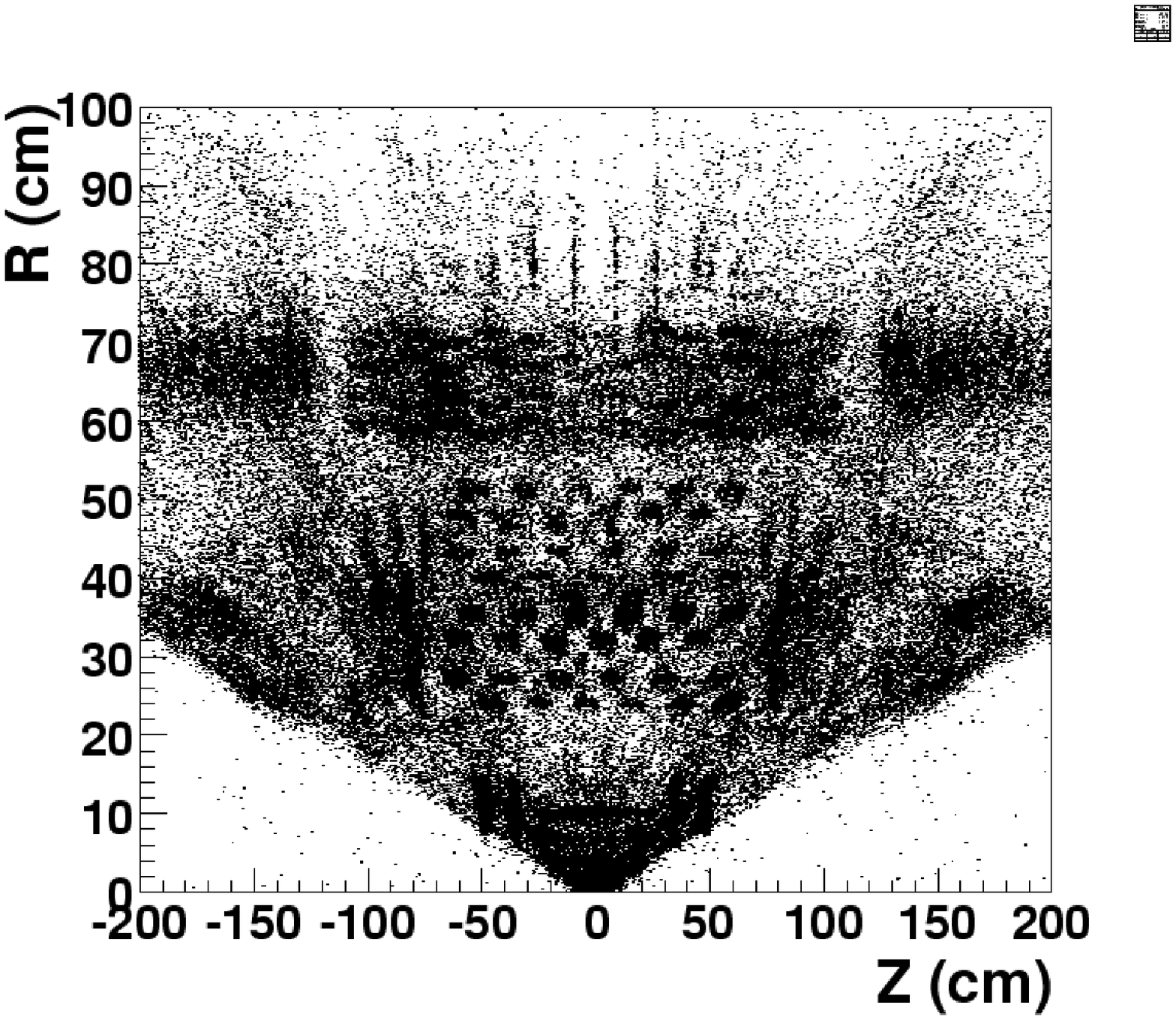}
\includegraphics{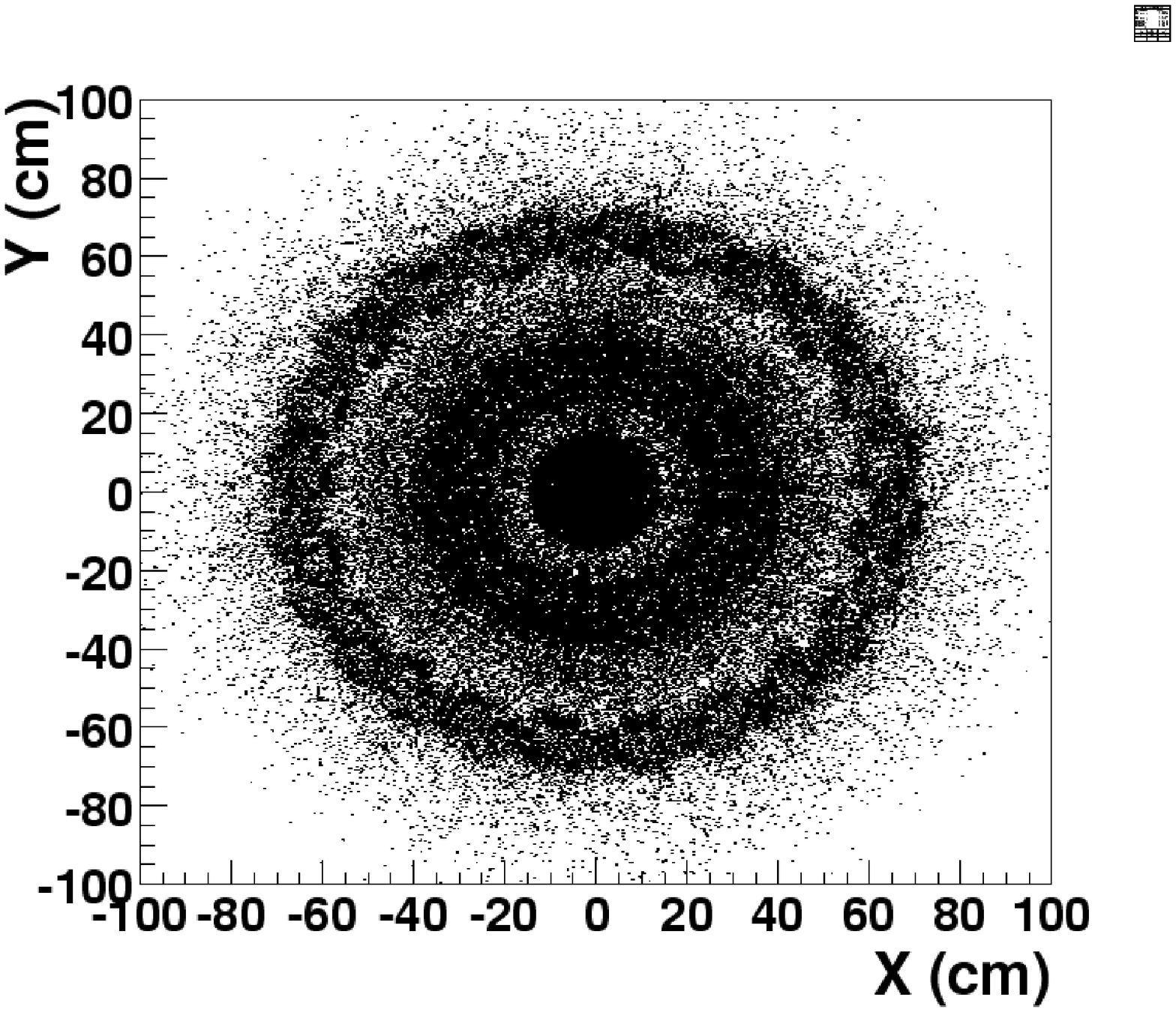}

\end{center}
\vspace{-2cm}
\caption{\small {Reconstructed converted photon vertices.}}
\label{Fig4}
\end{figure}

At this point it is important to check that the photon momentum measured from the
tracks matches the energy collected in the ECAL super-cluster. 
The ratio $P_T({\mathrm {tracks}})/E_T({\mathrm {SC}})$ is shown in Fig.~\ref{Fig3}
for signal and background (dark grey) from fake pairs. The fraction of fake pairs was measured in a
sample of Higgs-to-two-photon decays with low LHC luminosity (\mbox{10$^{33}$ cm$^{-2}$ s$^{-1}$}) pile-up events; it amounts
to about 5\%, easily reducible with a cut on $P_T({\mathrm {tracks}})/E_T({\mathrm {SC}})$.
Finally the position of the fitted conversion vertices is shown in Fig.~\ref{Fig4}. 
It is worth emphasizing that the results presented here were obtained with a non-final
simulation of the Tracker material and are likely to change in a future
update.

\section{Conclusions}
A baseline reconstruction method for converted photons in CMS has been described. It gives very
encouraging results.  This tracking method, designed specifically for
reconstruction of converted photons  has recently been ported to the 
new CMS Software environment (CMSSW) (Ref.~\cite{CMSSW}), where the final Tracker geometry description
is being finalized.

\section*{Acknowledgments}
I wish to thank all CMS colleagues who, over the years, contributed to the development of 
the ECAL clustering and the basic tracking tools used for this work.
A special thank is for R. Ruchti who constantly provides me with sound
advices. Finally I would like to acknowledge the US National Science Foundation
for financial support under grants PHY-0355340 and  PHY-0735119.

\end{document}